# Comprehensive Methodology for Sample Augmentation in EEG Biomarker Studies for Alzheimer's Risk Classification

## Authors:

Verónica, Henao Isaza[1,3]; David, Aguillon[2]; Carlos Andrés, Tobón Quintero[1,2]; Francisco, Lopera[2]; John Fredy, Ochoa Gómez[1,2,3].

1. Grupo Neuropsicología y Conducta, Universidad de Antioquia, School of Medicine. Medellín, Colombia.

2. Grupo de Neurociencias de Antioquia, Universidad de Antioquia, School of Medicine. Medellín, Colombia.

3. Semillero de Investigación NeuroCo, Universidad de Antioquia, School of Medicine & School of Engineering. Medellín, Colombia.

## Abstract

**Background:** Dementia, characterized by progressive cognitive decline, is a major global health challenge. Alzheimer's disease (AD) is the predominant type, accounting for approximately 70% of dementia cases worldwide. Electroencephalography (EEG)-derived measures have shown potential in identifying AD risk, but obtaining sufficiently large samples for reliable comparisons remains a challenge. **Objective:** This study implements a comprehensive methodology that integrates signal processing, data harmonization, and statistical techniques to increase sample size and improve the reliability of Alzheimer's disease risk classification models. **Methods:** We used a multi-step approach combining advanced EEG preprocessing, feature extraction, harmonization techniques, and propensity score matching (PSM) to optimize the balance between healthy non-carriers (HC) and asymptomatic E280A mutation Alzheimer's disease carriers (ACr). Data were harmonized across four databases, adjusting for site effects while preserving important covariate effects such as age and sex. PSM was applied at different ratios (2:1, 5:1, and 10:1) to explore the impact of sample size differences on model performance. The final dataset was subjected to machine learning analysis using decision trees, with cross-validation to ensure robust model performance. **Results:** Our analysis revealed significant improvements in classification accuracy when balancing sample sizes using PSM, with accuracy metrics ranging from 0.92 to 0.96 across different class distribution ratios. The refined data-driven approach enabled more accurate identification of individuals at risk for AD, even in populations with limited sample sizes. **Conclusion:** This comprehensive methodology highlights the potential of integrating data processing, harmonization, and statistical balancing techniques to improve the accuracy of Alzheimer's disease risk classification models, paving the way for the use in other neurodegenerative diseases.

## INTRODUCTION

The treatment of neurodegenerative diseases, particularly Alzheimer's disease (AD), is emerging as a major global health concern [1]. AD, which is characterized by progressive cognitive decline, imposes a significant burden on individuals and society at large, contributing to approximately 70% of all cases of dementia worldwide[2]. Its hallmark features include the accumulation of beta-amyloid (Aβ) plaques and hyperphosphorylated tau neurofibrillary tangles, leading to neurodegeneration over time [3]. While understanding of the underlying mechanisms has advanced in recent decades, early and accurate detection of AD remains a major obstacle in clinical practice[4].

Several biomarkers for Alzheimer's disease have been identified in the literature; however, the use of electroencephalography (EEG) biomarkers for reliable Alzheimer's disease risk classification remains relatively unexplored [5]. Traditional methods often face challenges in balancing healthy non-carrier subjects (HC) and asymptomatic E280A mutation Alzheimer's disease carriers (ACr) groups, as well as in accounting for demographic variables such as age and sex [6]. Despite the potential of EEG in distinguishing subjects in preclinical stages of AD, obtaining sufficiently large samples for reliable comparisons remains a significant challenge.

Recent studies have highlighted the importance of large-scale collaborations that emphasize the integration of diverse data sets [7], [8], [9]. The EEG-IP platform, as presented in the work of van Noordt et al.[10] serves as an exemplary model for successful integration of infant EEG data from multiple sites. By pooling longitudinal cohort studies, adhering to the Brain Imaging Data Structure (BIDS) EEG standard, and implementing a common signal processing pipeline, a standardized and integrated dataset was achieved[11]. This pioneering effort highlights both the successes and challenges encountered, particularly in addressing issues related to signal annotation, timing, and independent component analysis [12] during preprocessing.

Similarly, the work of Duncan et al. [13] introduces the Data Archive for the BRAIN Initiative (DABI), a dedicated platform designed to address the complexities of sharing human intracranial neurophysiology recordings and multimodal data. This initiative aligns with the overarching goal of creating specialized repositories capable of accommodating the unique features of complex and heterogeneous datasets, further validating the feasibility and importance of harmonizing data from diverse sources [14].

The main objective of this study is to develop a framework for exploring differences in pre-symptomatic subjects with scarce samples using advanced machine learning (ML) techniques. Using propensity score matching (PSM) techniques [15], we aim to optimize the balance between HC and ACr. Our study provides a detailed analysis of the results obtained, allowing us to identify significant patterns and differences in the distribution of relative performance between the groups of interest. Data visualization through figures and tables helps us to better understand the peculiarities of the EEGs of ACr compared HC.

This paper presents original research focused on an innovative workflow that integrates data from multiple sources and employs advanced data analysis techniques. By addressing the urgent need to improve the early and accurate detection of Alzheimer's disease, we aim to provide new insights that will drive the development of more effective approaches to AD risk classification.

## MATERIALS AND METHODS

### A. Subjects and EEG acquisition

The project consists of four databases, each with specific subject groups and data acquisition protocols.

1. **<u>UdeA 1 Database</u>:** Included 27 ACr Colombian kindred, 17 HC. *Acquisition Protocol*: EEG signals were recorded in resting states with closed eyes and open eyes for 5 minutes, using a Neuroscan amplifier and a 58-tin channel cap.
2. **<u>UdeA 2 Database:</u>** Included 22 ACr Colombian kindred, 12 HC. *Acquisition Protocol*: EEG recordings for 5 minutes were obtained from 64 electrodes with eyes closed, using a Neuroscan amplifier.
3. **<u>SRM Database:</u>** Included 31 HC. *Acquisition Protocol*: The data is recorded with a BioSemi ActiveTwo system, using 64 electrodes following the positional scheme of the extended 10-20 system (10-10). Resting-state EEG for 4 minutes.
4. **<u>CHBMP Database:</u>** Included 38 HC. *Acquisition Protocol*: Resting-state EEG for 10 minutes was recorded using a digital electroencephalograph system MEDICID with 64 and 128 electrodes.
5. **<u>All Databases:</u>** The pooled data set includes records of ACr and HC. We considered multi-center recordings of EEG studies with neurophysiological and neuropsychological information. Table 1. shows the general characteristics of the study datasets.

The information used consists of 237 to 173 records, divided into two groups: healthy non-carrier subjects (HC) and asymptomatic E280A mutation Alzheimer's disease carriers (ACr). Basic demographic information is available for almost all the subjects, presented in Table 1. which includes their respective averages and standard deviations.

**Insert here Table 1**

## B. EEG data pre-processing

The raw data underwent pre-processing using the pipeline proposed by Suarez et al.[16]. The standardized early-stage EEG (PREP) processing pipeline was applied (Fig 1), including signal detrending, robust referencing, and interpolation of bad channels. The Fast ICA algorithm obtained artifactual and neural ICA components after applying a high-pass filter. The records were segmented into 5-second epochs and subjected to wavelet-ICA for further artifact removal. At this step, individual Infomax ICA was applied. A low-pass filter was applied, and noisy epochs were detected and removed based on various criteria. The data was normalized to account for variability introduced by hair, scalp, and skull. Spectral, connectivity, and amplitude modulation features were extracted.

**Insert here Fig 1**

The Ochoa-Gómez, J. F., et al. [17] gICA methodology, outlined in the latter study, served as a robust foundation for extracting reproducible neuronal components from resting-state electroencephalographic data. This methodological framework ensures the reliability and reproducibility of the independent components, which were used as spatial filters for extracting the signals analyzed in this study.

Building on this, we applied machine learning models to harmonize features across diverse cohorts, focusing on Alzheimer's disease and PSEN1-E280A mutation carriers (ACr).

Additionally, feature extraction involved assessing several key metrics:

Relative Power: This measure evaluates the proportion of a signal's power relative to a reference, providing insight into neural activity in different brain regions [18].

Entropy: A measure of uncertainty or disorder in a dataset, used to characterize the complexity and regularity of brain activity patterns [19].

Coherence: A measure of the consistency or synchronization between signals at different frequencies, indicating functional communication between brain regions [20].

Cross Frequency: Examines the relationship between oscillations in different frequency bands, providing insight into the organization and integration of neural activity [21].

Synchronization Likelihood: Assesses the likelihood that two signals are synchronized in time, reflecting functional connectivity between brain regions [22] .

### C. Harmonization

To ensure harmonization across different datasets, we employed the neuroHarmonize package [23]. This tool, which extends the functionality of neuroCombat [24], uses the ComBat algorithm for correcting multi-site data. The data matrix and covariate matrix were prepared and harmonized, controlling for site effects while preserving covariate effects. This step was crucial for reducing inter-subject variability and improving the consistency of the data, thereby enhancing the reliability of subsequent analyses.

The propensity score matching (PSM) process used logistic regression to match individuals from the ACr and HC subjects based on their similarity in pretreatment covariates such as sex and age [15]. The propensity score calculated the probability of an individual being a gene carrier based on these characteristics. Subjects with lower propensity scores were then removed to achieve the desired proportion of subjects in the treatment group. PSM was used to ensure comparability between the healthy non-carrier (HC) and asymptomatic E280A mutation Alzheimer's disease carrier (ACr) groups in the study. This method aims to balance covariates between treatment and control groups in observational studies, thereby providing more valid comparisons and reducing bias due to non-random treatment allocation.

This logistic regression-based propensity score matching ensures that confounding variables are controlled, leading to more comparable and homogeneous groups. By refining the matching process, differences observed in the EEG data can be more accurately attributed to the gene carrier status rather than initial characteristic differences.

### D. Subject Ratios and Age-Sex Matching

We employed Propensity Score Matching (PSM) at varying ratios (2:1, 5:1, and 10:1) to assess the impact of sample size disparities between healthy non-carrier subjects (HC) and asymptomatic E280A mutation Alzheimer's disease carriers (ACr) on model performance. These ratios were chosen to explore how increasing the size of the HC cohort relative to ACr influences the robustness and generalizability of our findings. Through PSM, we aimed to achieve balance in demographic variables such as age and sex, thereby mitigating biases associated with non-random treatment assignment and improving the comparability between groups.

In Fig 2 the data input for both HC and ACr groups were combined across all cohorts (UdeA 1, UdeA 2, SRM, CHBMP) and subjected to propensity score matching (PSM). This process resulted in matched datasets with two, five, and ten times as many HC as ACr.

**Insert here Fig 2**

## E. Model selection.

The data was organized within dataframes, where each row corresponded to a record, and the columns represented specific features. For the implementation and validation of the model, an 80-20 train-test split was applied, ensuring that class proportions were maintained to provide an unbiased model evaluation on unseen data.

The cross-validation process employed a pre-fitted estimator for predicting outcomes, using ten-fold iterations for training and evaluation. Performance scores were averaged across folds to ensure robust model assessment.

The pseudocode below outlines the feature selection and model evaluation process, which highlights the steps taken to identify the most relevant features and optimize model precision:

**Feature Selection and Model Evaluation Process**

**Input:** Harmonized data D, feature set F, accuracy threshold T
**Output:** Final model M, confusion matrix C

```
Load harmonized data: D = load_data()
Initialize selected feature list: S = []
for each feature f in F do
    Train model with current feature: M_f = train_model(D, f)
    Evaluate model accuracy: A_f = evaluate_model(M_f)
    if A_f ≥ T then
        Add feature to selected list: S.append(f)
    end if
end for
Assign weights to selected features: W = assign_weights(S)
Retrain model with selected features: M = train_model(D, S, W)
Generate confusion matrix: C = generate_confusion_matrix(M)
Return final model M and confusion matrix C
```

Following this process, the features that contributed most to improving model accuracy were selected. The decision tree model, which achieved the highest precision, was subsequently employed to generate a confusion matrix, effectively evaluating the model's ability to classify individuals at risk of Alzheimer's disease.

Decision trees were chosen for their intrinsic feature selection capabilities, enabling a deeper analysis of feature importance. This process allowed us to assess how individual features influenced the classification of patients, providing insights into both their individual and collective impact on model performance.

Additionally, the effect size of each selected feature was evaluated using Cohen's d. This standardized measure quantified the difference between ACr and HC groups for each specific metric, offering a comprehensive understanding of each feature's contribution to group differentiation.

## RESULTS

Our study utilized an extensive dataset to develop a model capable of identifying statistical differences and distinguishing between ACr and HC across different cohorts. This comprehensive approach allowed us to provide a more nuanced and precise understanding of the implications of the PSEN1-E280A mutation in varied carrier contexts. The findings have significant clinical and research implications, offering potential advancements in early diagnosis and targeted interventions for Alzheimer's disease. By integrating data from diverse sources, our model enhances the robustness and generalizability of EEG-based classification, contributing valuable insights to the field of Alzheimer's research.

### A. Preprocessing

By implementing the previously validated processing pipeline from the study of Henao Isaza et al. [25]. The relative power spectral density was computed by analyzing EEG frequency bands, including delta (1.5-6 Hz), theta (6-8.5 Hz), alpha 1 (8.5-10.5 Hz), alpha 2 (10.5-12.5 Hz), beta 1 (12.5-18.5 Hz), beta 2 (18.5-21 Hz), beta 3 (21-30 Hz), and gamma (30-45 Hz) [26].

**Insert here Fig 3**

In Fig 3, The graph illustrates the comparison of components (ICs) of the Beta3 band across four cohorts of interest. Each boxplot represents the distribution of relative power values for a specific component (C) within different groups and databases.

For component C1, the median power values range from 0.05 to 0.1 across all subjects in the four cohorts. However, outliers are observed for the SRM and CHBMP HC groups, with the CHBMP group exhibiting an outlier reaching a value of 0.30 for relative power. In addition, only the UdeA1 cohort shows an outlier for the ACr group.

For Component C2, the median power values across subjects are even closer, with outliers detected in the CHBMP and UdeA2 cohorts. Notably, in the UdeA2 cohort, outliers are present in both the HC and ACr groups. Component C3 shows greater variability in median power values, with outliers in the UdeA2 cohort. Components C4, C5, C6 and C8 show a higher prevalence of outliers, especially in the HC group. In components C7 and C9, the median power values range from 0.075 to 0.150, with outliers observed in the SRM, UdeA1, and UdeA2 cohorts for the ACr group.

## B. Matching between subjects using propensity score matching (PSM)

By estimating the propensity score, which is the probability of receiving a particular treatment given observed covariates, individuals in the treatment group can be matched with individuals in the control group who have similar propensity scores.

Fig 4 shows the range of common support by treatment status. The propensity score is plotted on the x-axis and the frequency is plotted on the y-axis. The common support functions are smooth, and the algorithm's balancing property is satisfied. Beneficiaries outside the range of shared support were dropped in the linear regression models for each case (2:1, 5:1, 10:1).

**Insert here Fig 4**

## C. Feature selection

It is important to select relevant features that capture meaningful aspects of brain activity before training a model for EEG analysis. The features used are presented in Fig 1, including the components from [17] and the frequency bands described in the processing, some of which are shown in Fig 5. These features were selected based on their relevance to capturing meaningful aspects of brain activity for EEG analysis. From nearly 967 initial features (Fig 1), the model first removed those with the highest correlations. The decision tree algorithm then identified the 100 most important features for inclusion (Model Selection). Cohen's d values were utilized to quantify the magnitude of differences between groups ACr and HC for each specific metric.

Fig 5 illustrates the distribution of effect sizes (Cohen's d) for selected metrics in EEG analysis, highlighting the observed differences between ACr and HC groups. Cohen's d provides standardized measures of effect size: Small (0.2): Indicates a small difference. Medium (0.5): Represents a moderate difference. Large (0.8): Indicates a substantial difference. Very Large (1.2+): Represents a significant difference.

Relative power analysis revealed a Cohen's d of 0.52 for component C5 in the Beta3 band, indicating moderate differences between ACr and HC groups. Similarly, cross-frequency analysis in component C1 showed a Cohen's d of 1.03 for the modulated Beta3 and Beta3 bands, signifying substantial differences. Coherence analysis with component C4 in the Theta band exhibited a Cohen's d of -1.56, while synchronization likelihood in component C1 in the Alpha1 band had a Cohen's d of -0.97, indicating significant differences in opposite directions.

In addition, these features were evaluated using Cohen's d to assess effect size, comparing group differences across three proportions (n1, n2, and n3). An initial model training was conducted on the n1 dataset, followed by incremental model adjustments on n2 and n3 datasets. This iterative approach ensured comprehensive model refinement and evaluation across different dataset compositions.

**Insert here Fig 5**

D. **Train and Test** For the 2:1 ratio, the training curve starts at around 0.5 scores and gradually increases, stabilizing at approximately 0.8 after 30 samples. In contrast, the validation curve begins around 0.4 scores, exhibits a slightly steeper rise, stabilizes at about 0.3 after 30 samples, and then climbs to about 0.87 after 130 samples, peaking finally at approximately 0.91.

In the case of the 5:1 ratio, the training curve starts near a score of 0.99, dips slightly around 50 samples, and rises again to about 1.0 by 100 samples. Meanwhile, the validation curve begins at a score of 0.83, experiences a slight dip, rises again around 40 samples, and gradually increases to about 0.98.

Lastly, for the 10:1 ratio, the training curve starts near 0.91 scores, rises rapidly to nearly 0.95, dips slightly around 60 samples, and then rises to 0.97. The validation curve for this ratio begins at approximately 0.9 and grows steadily to 0.96, demonstrating the most stable performance among all ratios.

**Insert here Fig 6**

### E. Confusion matrix

**Insert here Table 2**

Table ***2*** presents the results of the computer precision function, which likely calculates the precision (and possibly other metrics) on the test dataset. By comparing the model's predictions with the true labels, this result is crucial as it evaluates the model's performance on unseen data, providing a more realistic estimate of how the model will perform on future data.

The figure shows three confusion matrices for class distribution ratios of 2 to 1, 5 to 1, and 10 to 1, providing a visual representation of the computer precision results. In these matrices, TP stands for true positives, FP for false positives, FN for false negatives, and TN for true negatives.

For the 2:1 ratio, the results are TP = 31, FP = 1, FN = 3, and TN = 13. For the 5:1 ratio, the results are TP = 32, FP = 0, FN = 1, and TN = 5. For the 10:1 ratio, the results are TP = 32, FP = 0, FN = 2, and TN = 1. These matrices provide insight into the performance of the model at different class distribution ratios, highlighting the differences in correct and incorrect predictions.

**Insert here Fig 7**

## DISCUSSION

Integrating data from multiple cohorts is fundamental to improving the robustness and generalizability of EEG-based AD detection models [27]. Our study took a comprehensive approach by combining data from multiple sources. This highlights the importance of considering the diversity of data in Alzheimer's disease research to ensure that findings apply to a wide range of populations and clinical scenarios [28].

Our discussion focuses on how PSM and gICA improve the accuracy and reliability of our model, highlighting the importance of considering different methodological approaches to address unique challenges in early AD detection. This comparison underscores the need for comprehensive and multifaceted strategies to improve the accuracy and applicability of EEG-based AD detection models [29].

The figure in Fig 6 illustrates different patterns in the learning curves for different class distribution ratios. For the 2:1 ratio, both the training and validation curves show a gradual increase in accuracy, with validation slightly outperforming training after a certain point, reaching a validation result of 91%. For the 5:1 ratio, the validation curve had a slight initial dip, ending at 98%. At the 10:1 ratio, both curves quickly reach near 96% accuracy, with the validation curve showing the greatest stability and steady growth in accuracy over time.

The results from the performance evaluation table (Table 2) reveal varying metrics across different class distribution ratios. For the 2:1 ratio, the accuracy is 91%, precision is 91%, recall is 97%, F1-score is 94%, and AUC is 92%. For the 5:1 ratio, accuracy improves to 98%, precision to 97%, recall to 100%, F1-score to 98%, and AUC significantly increases to 99%. The 10:1 ratio shows accuracy at 96%, precision at 97%, recall at 100%, F1-score at 98%, and AUC at 93%.

The notably high AUC of 99% in the 5:1 ratio suggests that the balance in the dataset for this particular ratio may facilitate optimal model performance. This configuration likely provides a better equilibrium between subjects and controls, leading to more effective learning and classification by the algorithm. This balance may allow the model to better discriminate between classes, resulting in an exceptionally high AUC. This is particularly significant given that earlier studies have reported varying results. For instance, recent research has reported AUC values ranging from 0.962 to 1.0 [30], while earlier studies have shown an AUC value of 0.85 [31].

This variability in AUC outcomes can be attributed to differences in the data used (such as sample size, data quality, and specific population characteristics) and the analytical methods employed (including the types of features utilized and data preprocessing techniques). Each study may employ different datasets and methodologies, which can significantly affect the resulting AUC values.

Previous research by García-Pretelt et al. [32] applies machine learning to classify individuals at risk for Alzheimer's disease using resting-state EEG, achieving an impressive accuracy of up to 83% using spatial filters obtained from a gICA approach. Additionally, the study by Francisco Gerson A de Meneses et al. [33] provides insights by using convolutional neural networks (CNNs) to classify neurological diseases based on cortical topographies. The remarkable performance of SqueezeNet, with accuracies of 88.89% for Parkinson's disease, 75.70% for depression, and 72.10% for bipolar disorder, highlights the potential of advanced machine learning techniques in the classification of neurological conditions, which complements our study's focus on ICA configurations.

Caroline L. Alves et al. [34] focuses primarily on harmonizing EEG data from different cohorts, with accuracies of 98 and 99% using CNNs in Parkinson's disease. On the other hand, the present study takes a broader approach, integrating and analyzing data from different sources to improve Alzheimer's risk classification. This strategy has allowed us to gain a more complete and accurate understanding of the impact of the PSEN1-E280A mutation in different carrier contexts. This underscores the importance of considering data diversity in AD research to ensure model robustness and generalizability.

The results of Gerson et al. [33] serve as a reference point, demonstrating the successful application of similar algorithms with EEG data in various diseases, thus reinforcing the robustness of our methods. This, together with other papers[35], highlights the importance of our findings in the clinical context and underlines the implications for future research and clinical applications. However, our study goes further by providing a more comprehensive understanding of how integrating data from different cohorts can improve the accuracy and generalizability of EEG-based AD detection models. This underscores the need to continue to explore innovative and collaborative approaches to address challenges in Alzheimer's diagnosis and treatment.

On the other hand, the components obtained from the reproducibility approach reported in the study by Ochoa-Gómez et al. [17] provide a variety of metrics for classification, in line with the proposal of Prado et al. [36]. This study highlights the need for systematic harmonization in EEG connectivity studies, addressing critical sources of variability and suggesting a composite metric strategy to improve replicability in multicenter studies.

While our study is promising, it's important to acknowledge its limitations. Despite using a comprehensive dataset, we must question its representativeness and potential biases, especially with similar cohorts such as UdeA1 and UdeA2. Generalizing our findings beyond these specific cohorts may be challenging, highlighting the need for validation in diverse populations and increasing cohort diversity [37]. In addition, despite the use of advanced machine learning techniques such as EEG-based classification, the complexity of the models may hinder full interpretation. Finally, potential biases, measurement errors, and underlying assumptions in our analysis models warrant careful consideration and further discussion, particularly regarding model selection and the preference for SVM in the literature.

## CONCLUSIONS

Our study represents a significant step forward in the field of computational neuroscience, particularly in the area of Alzheimer's disease detection using EEG data. By harmonizing data from multiple cohorts and applying advanced machine learning techniques, we have developed a robust model capable of accurately discriminating between healthy non-carrier subjects (HC) and asymptomatic E280A mutation Alzheimer's disease carriers (ACr).

Our results highlight the importance of integrating data from multiple sources to improve the generalizability and applicability of disease detection models. The successful application of techniques such as propensity score matching (PSM) and group independent component analysis (gICA) highlights the effectiveness of comprehensive approaches in improving model accuracy and reliability.

While our study presents promising results, it is imperative to acknowledge its limitations. The representativeness of our dataset and potential biases inherent in our sample must be carefully considered. Furthermore, the complexity of our machine learning model requires thorough sensitivity analyses and cross-validation to ensure its stability and robustness.

In conclusion, our work contributes valuable insights to the burgeoning field of computational neuroscience, provides a refined understanding of the impact of the PSEN1-E280A mutation, and paves the way for more accurate and early detection methods for Alzheimer's disease. We believe that our findings have significant clinical and research implications and represent a critical step in addressing the challenges posed by neurodegenerative diseases in our aging population.

## ACKOWLEDGEMENTS

The authors acknowledge the support provided by Comité para el Desarrollo de la Investigación - CODI Universidad de Antioquia, through the project "Cambios en los patrones del electroencefalograma cuantitativo (reactividad alfa, theta y su índice) en reposo y tareas de memoria, en el seguimiento longitudinal de pacientes con riesgo genético para Enfermedad de Alzheimer Temprano", identified with the code 2017-16371.

## DATA AVAILABILITY

The data from the subjects are categorized into public and private datasets. The **CHBMP** and **SRM** databases are publicly available at the following links:

- CHBMP: https://chbmp-open.loris.ca/
- SRM: https://openneuro.org/datasets/ds003775/versions/1.0.0

The datasets named **UdeA1** and **UdeA2** are private; however, researchers are encouraged to reach out for potential collaborations. The codes used for analysis are publicly available in the following repositories:

- eeg_harmonization: https://github.com/GRUNECO/eeg_harmonization
- Data_analysis_ML_Harmonization_Proyect: https://github.com/GRUNECO/Data_analysis_ML_Harmonization_Proyect

## DECLARATION OF COMPETING INTEREST

The authors have no conflicts of interest to declare.

All subjects signed an informed consent approved by the ethics committee of the Instituto de Investigaciones Médicas of Universidad de Antioquia (Act no. 010, code F-017-00).

## AUTHOR CONTRIBUTIONS

**A**: conception and design; **B**: data curation; **C**: data analysis; **D**: interpretation of data; **E**: writing the manuscript draft; **F**: review and approval of the final manuscript.

Verónica Henao Isaza: A, B, C, D, E, F

Carlos Andrés, Tobón Quintero: F

David, Aguillon: F

Francisco, Lopera: F

John Fredy, Ochoa Gómez: A, D, E, F

**Table 1**. Summary characteristics of the databases

| | Database | Group | Count | Age (Mean ± SD) | Sex (F/M) |
|---|---|---|---|---|---|
| 2:1 | CHBMP | HC | 38 | 27.63 ± 6.67 | 13/25 |
| | SRM | HC | 31 | 30.77 ± 5.21 | 19/12 |
| | UdeA1 | ACr | 68 | 35.81 ± 4.36 | 49/19 |
| | UdeA1 | HC | 77 | 30.45 ± 4.81 | 47/30 |
| | UdeA2 | ACr | 11 | 33.45 ± 3.64 | 9/2 |
| | UdeA2 | HC | 12 | 31.42 ± 7.15 | 10/2 |
| | **Total** | | **237** | | **147/90** |
| 5:1 | CHBMP | HC | 38 | 27.63 ± 6.67 | 13/25 |
| | SRM | HC | 31 | 30.77 ± 5.21 | 19/12 |
| | UdeA1 | ACr | 30 | 39.78 ± 2.85 | 21/9 |
| | UdeA1 | HC | 77 | 30.45 ± 4.81 | 47/30 |
| | UdeA2 | ACr | 1 | 43.0 ± nan | 1/0 |
| | UdeA2 | HC | 12 | 31.42 ± 7.15 | 10/2 |
| | **Total** | | **189** | | **111/78** |
| 10:1 | CHBMP | HC | 38 | 27.63 ± 6.67 | 13/25 |
| | SRM | HC | 31 | 30.77 ± 5.21 | 19/12 |
| | UdeA1 | ACr | 14 | 41.86 ± 2.35 | 12/2 |
| | UdeA1 | HC | 77 | 30.45 ± 4.81 | 47/30 |
| | UdeA2 | ACr | 1 | 43.0 ± nan | 1/0 |
| | UdeA2 | HC | 12 | 31.42 ± 7.15 | 10/2 |
| | **Total** | | **173** | | **102/71** |

*HC: Healthy non-carrier subjects. ACr: Asymptomatic E280A mutation Alzheimer's disease carriers.*

**Table 2.** Evaluation of Model Performance Using Computer Precision

| 2:1 | 5:1 | 10:1 |
|---|---|---|
| **Accuracy: 91%**<br>**Precision: 91%**<br>**Recall: 97%**<br>**F1-score: 94%**<br>**AUC: 92%** | **Accuracy: 98%**<br>**Precision: 97%**<br>**Recall: 100%**<br>**F1-score: 98%**<br>**AUC: 99%** | **Accuracy: 96%**<br>**Precision: 97%**<br>**Recall: 100%**<br>**F1-score: 98%**<br>**AUC: 93%** |

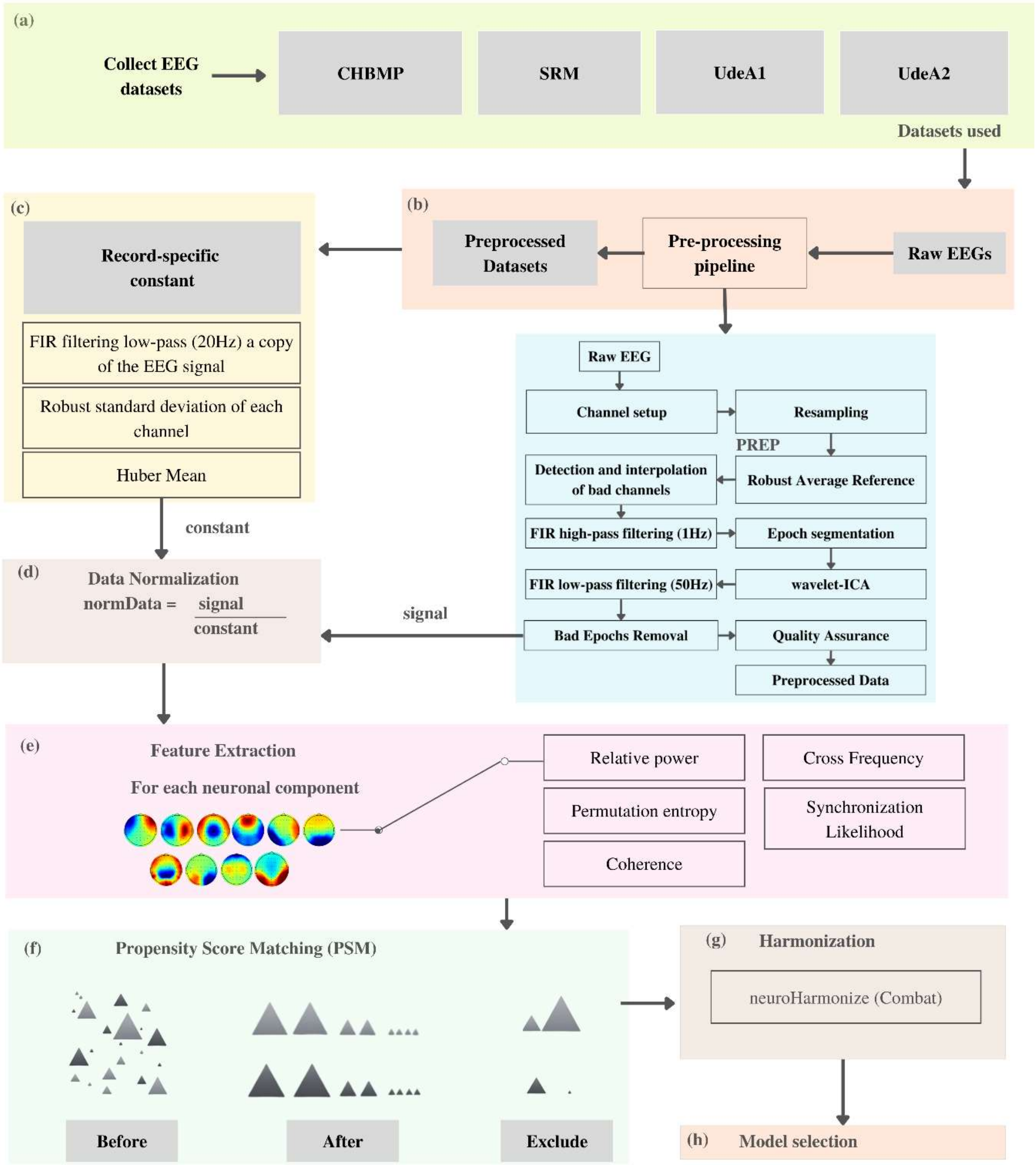

**Fig 1** The figure illustrates the step-by-step pipeline used for pre-processing EEG signals to enhance test-retest reliability, as outlined in the diagram of the EEG Signal Pre-processing Pipeline [16]. In this diagram, additional steps, including (d)normalization [36], (e)feature extraction, and (g)harmonization, are incorporated to further refine the EEG signal processing procedure. From the 5 metrics, 8 bands, 9 components, and demographic variables, a total of 967 features are generated.

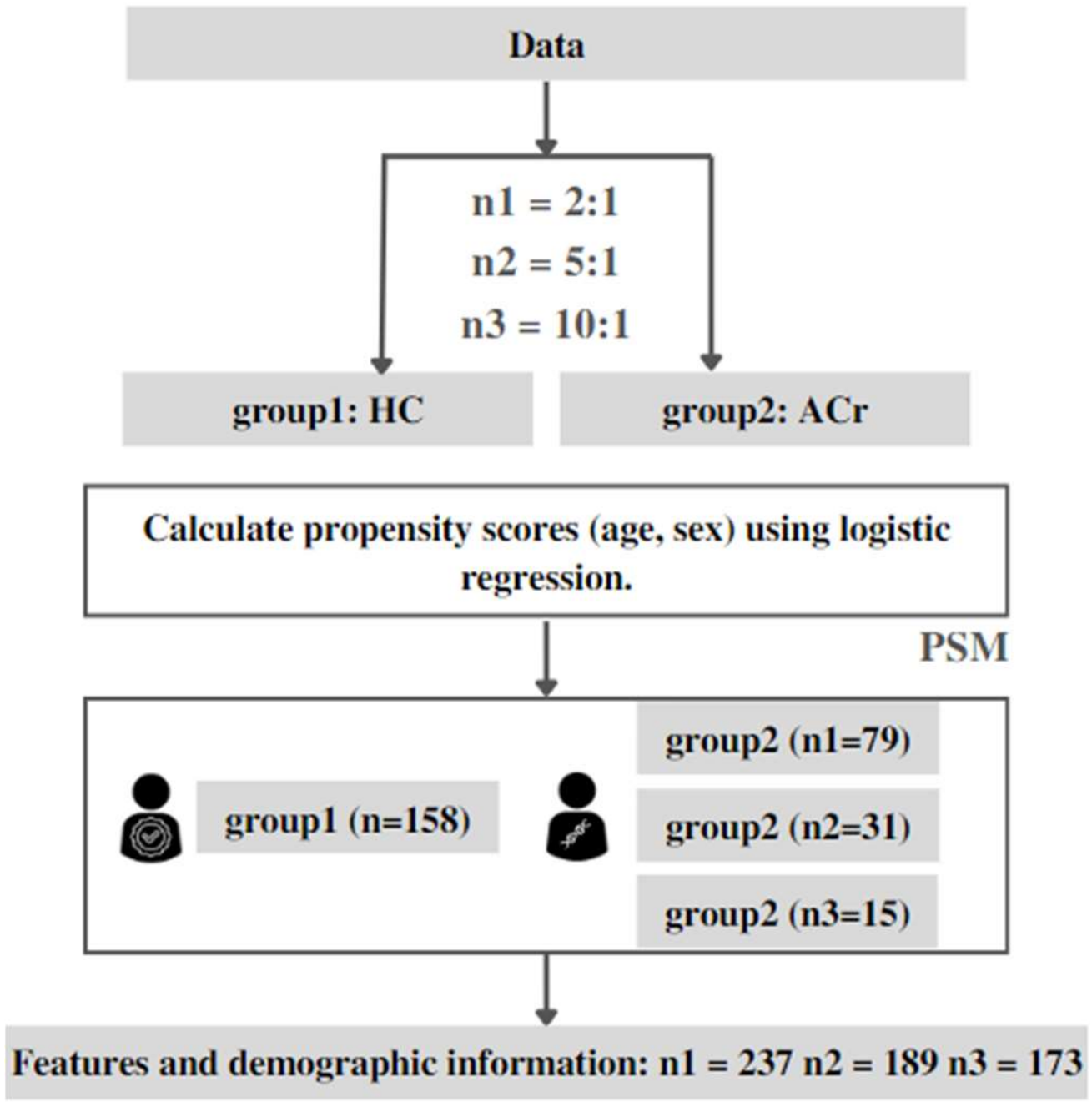

**Fig 2** 457 age- and sex-matched records for group 1, healthy non-carrier subjects (HC), and group 2, asymptomatic E280A mutation Alzheimer's disease carriers (ACr), were utilized using the PSM function at 2:1, 5:1, and 10:1 ratio. This process resulted in ACr groups comprising 79, 31, and 15 subjects, respectively, while the HC groups consisted of 158 subjects each. Consequently, the total number of subjects in each dataset is 237, 189, and 173, respectively.

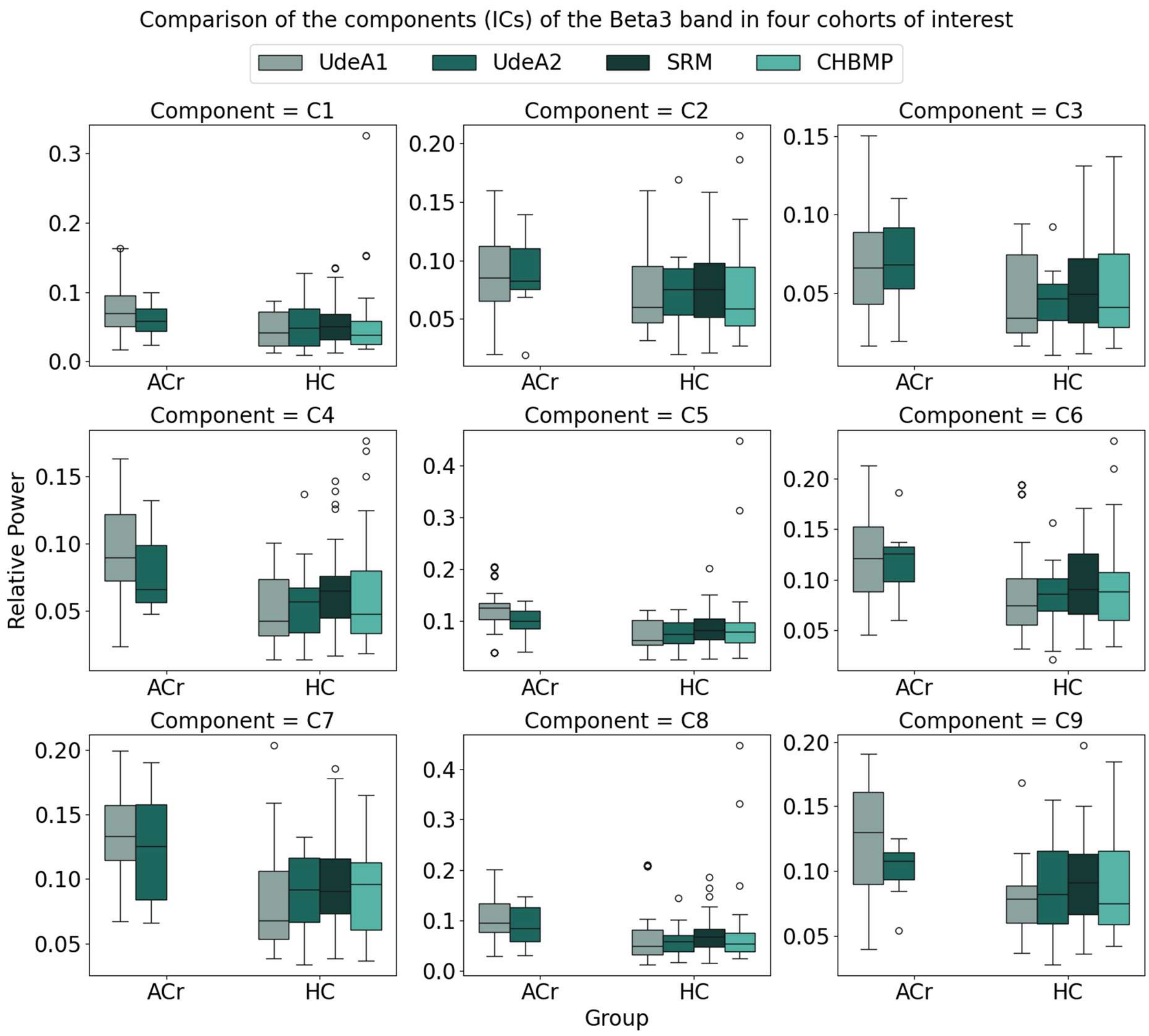

**Fig 3** Comparative analysis of relative Beta 3 performance among four cohorts.

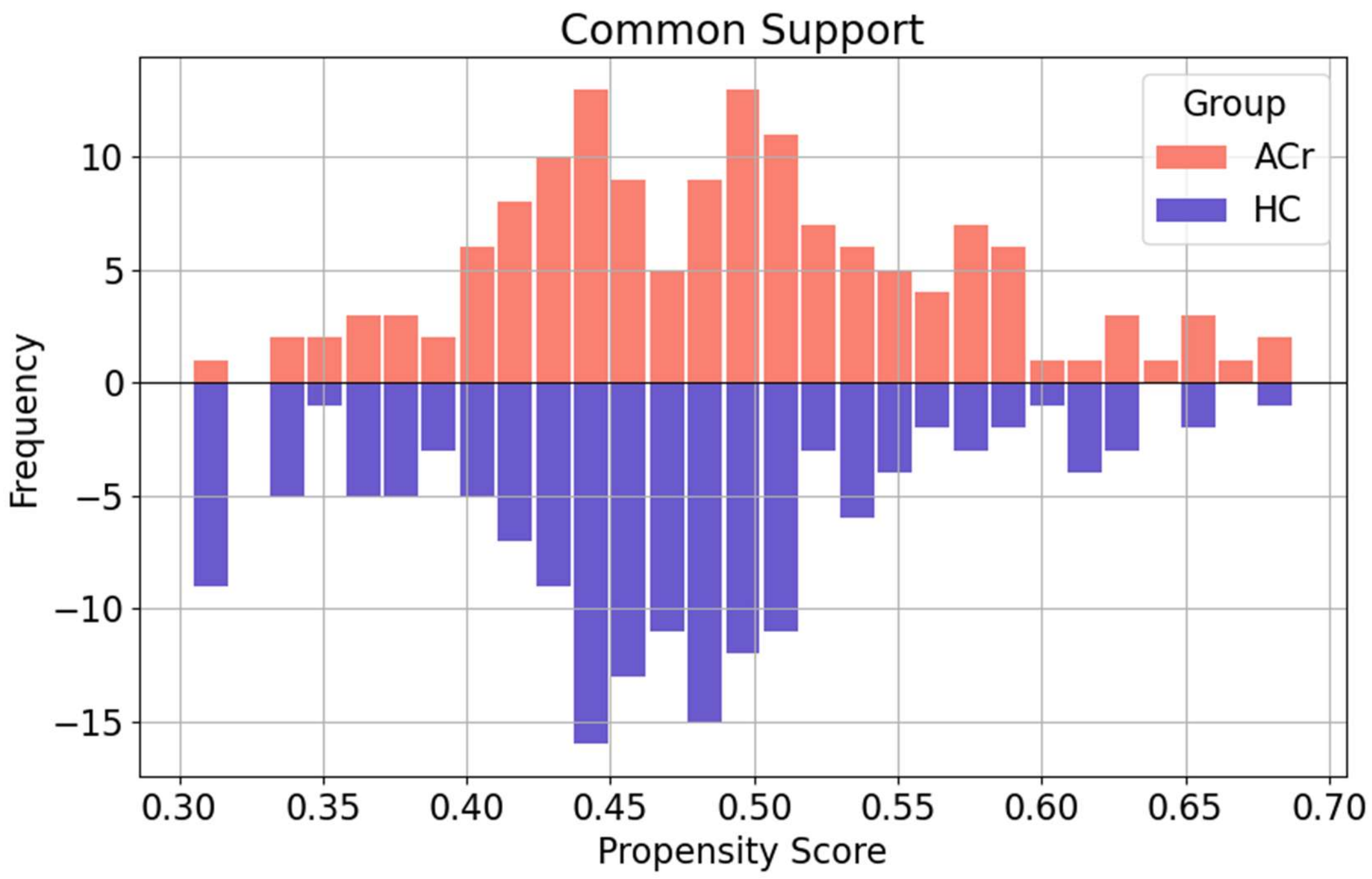

**Fig 4** Propensity score matching: area of common support.

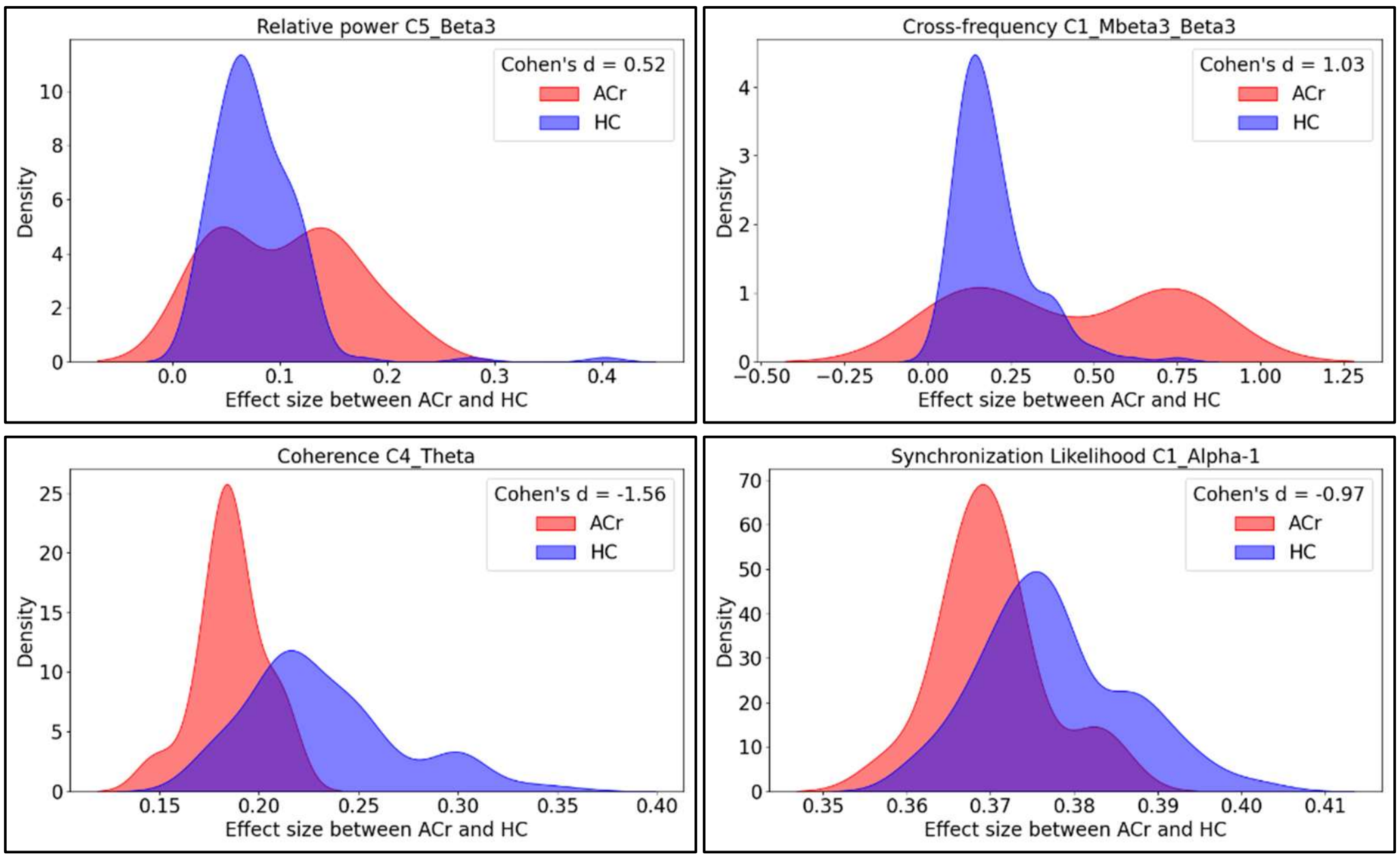

**Fig 5** Distribution of effect size (Cohen's d) for selected metrics in EEG analysis. The bars depict the magnitude of the effect between ACr and HC groups for different metrics, highlighting the observed differences.

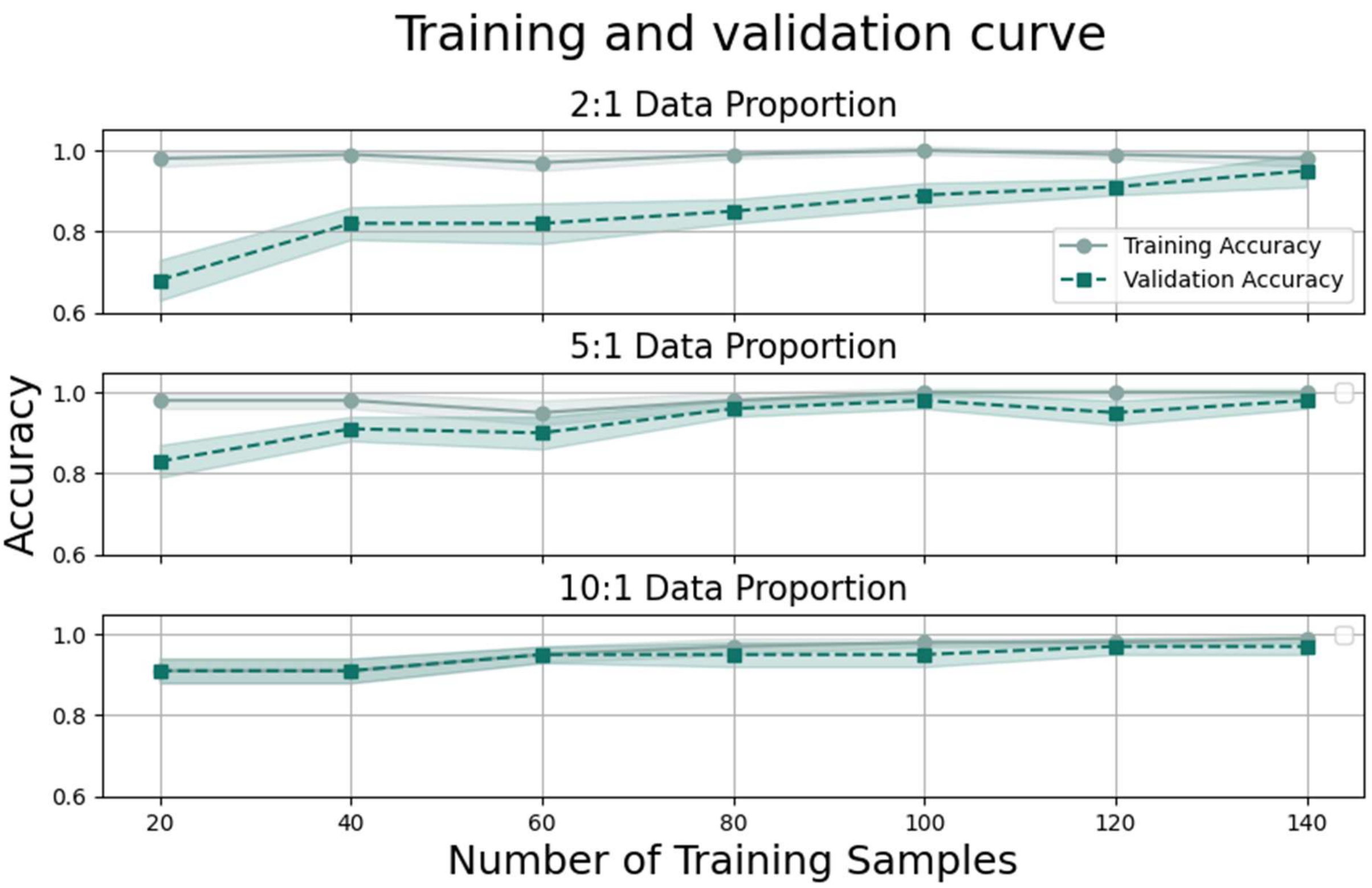

**Fig 6** Comparison of accuracy between learning curves on 2:1, 5:1, and 10:1 with different ratio configurations.

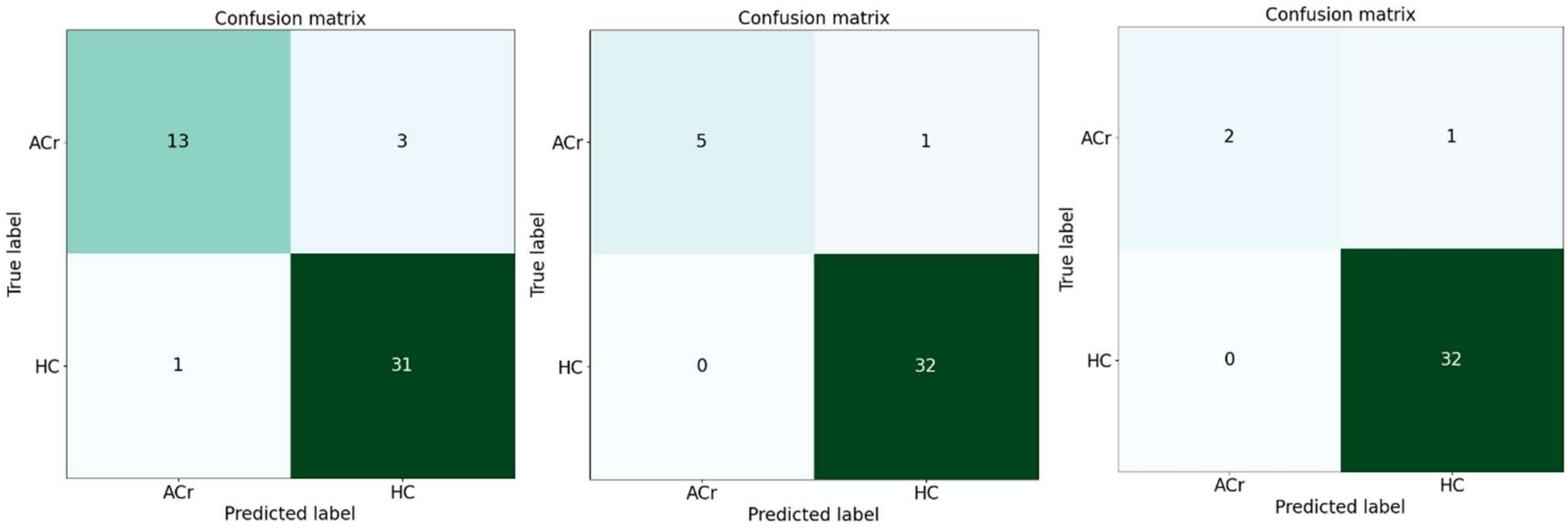

**Fig 7.** Comparison of confusion matrices for different ratios: Evaluated on 20% of total data. HC: healthy non-carrier subjects, and ACr: asymptomatic E280A mutation Alzheimer's disease carriers.